\documentclass[11pt]{article}
\usepackage{graphicx,wrapfig,color}
\usepackage{subfig}
\usepackage{setspace}
\usepackage{epsfig}
\usepackage{graphicx}
\usepackage{multicol}
\usepackage{amsfonts}
\usepackage{latexsym}
\usepackage{amsmath,amssymb}
\usepackage{enumerate}
\usepackage{hyperref}
\hypersetup{
    colorlinks=flase,
    linkcolor=black,
    filecolor=magenta,      
    urlcolor=blue,
}
\usepackage{cite}
 \usepackage{url} 
\usepackage{lipsum}
\usepackage{mwe}
\usepackage{subfig}
\newcommand{\ben}{\begin{enumerate}}
\newcommand{\een}{\end{enumerate}}
\newcommand{\spa}{\phantom{s}}
\newcommand{\bea}{\begin{eqnarray}}
\newcommand{\eea}{\end{eqnarray}}
\newcommand{\be}{\begin{equation}}
\def\bel#1{\begin{equation} \label{#1}}
\newcommand{\ee}{\end{equation}}
\newcommand{\bi}{\begin{itemize}}
\newcommand{\ei}{\end{itemize}}
\newcommand{\ba}{\begin{align}}
\newcommand{\ea}{\end{align}}

\def\bel#1{\begin{equation} \label{#1}}

\def\be{\begin{equation}}
\def\ee{\end{equation}}
\def\bea{\begin{eqnarray}}
\def\eea{\end{eqnarray}}

\def\spa{\phantom{a}}
\def\pref#1{(\ref{#1})}
\def\hf{\frac12}

\usepackage{color}
\definecolor{cblue}{RGB}{100,5,255}
\definecolor{cred}{RGB}{255,50,40} 
\definecolor{cgreen}{RGB}{40,255,40} 
\definecolor{corange}{RGB}{250,200,40}

\begin{document}

\begin{titlepage}
\vskip 1 cm
\begin{center}
{\large \bf{Post-inflationary Scalar Tensor Cosmology and Inflationary Parameters}}
\vskip 1.5cm  
{ 
{\sc{Anshuman Maharana}}$^{\dagger}$  and { \sc{Ivonne Zavala}}$^{*}$
\let\thefootnote\relax\footnotetext{ \hspace{-0.5cm} E-mail: {$\mathtt{anshumanmaharana@hri.res.in, e.i.zavalacarrasco@swansea.ac.uk} $}}
}
\vskip 0.9 cm

{\textsl{
$^{\dagger}$Harish Chandra Research Institute, \\
HBNI, Chattnag Road, Jhunsi,\\
Allahabad -  211019, India\\}
}

\vskip 0.6 cm

{\textsl{$^{*}$Department of Physics \\ 
Swansea University, Singleton Park, \\ 
Swansea, SA2 8PP, UK}
}

\end{center}

\vskip 0.6cm

\begin{abstract}
\vskip 0.5 cm

\noindent    Scalar fields provide attractive modifications of pre-BBN cosmology, which  have interesting  implications for dark matter abundances. 
We analyse the effect of these modifications on the number of e-foldings between horizon exit of CMB modes and the end of inflation $(N_k)$, and examine the consequences for inflationary predictions of various models. We find significant effects in the predictions in the $(n_s, r)$ plane. For a large part of the parameter space, the shift in $N_k$ is positive; this ameliorates the tension of  $m^{2} \varphi^{2}$ and natural inflation with data.
\end{abstract}

\vspace{3.0cm}

\end{titlepage}
\pagestyle{plain}
\setcounter{page}{1}
\newcounter{bean}
\baselineskip18pt
%



 \section{Introduction}
 
     Cosmological observations  are the source of some of the biggest challenges in Physics today. Only
$5 \%$ of the energy density of the universe is baryonic matter, the remaining is dark matter $(27 \%)$  and dark energy $(68 \%)$. 
The Cosmic Microwave Background (CMB) and galaxy distributions exhibit correlations even at super-horizon scales, with distinct patterns in the inhomogeneities.  The
$\Lambda CDM$ model together with an inflationary epoch in the early universe  has become the standard framework used to address these observations.

     In this setting, a  popular scenario to address dark matter abundance is the  Weakly Interacting Massive Particle (WIMP) paradigm. At early times, dark matter was in thermal equilibrium with the standard model degrees of freedom. As the temperature of the universe decreased, the dark matter
 interaction rate fell down and its number density became a constant in the comoving frame -- the present dark matter abundance is this relic. The 
 required value of the dark matter abundance is obtained for $  \langle \sigma_{\rm th} v_{\rm rel} \rangle \simeq 3 \times 10^{-26} \, \text{cm}^3 \, \text{sec}^{-1}$, where  $v_{\rm rel}$ is the Moller velocity and a thermal average is taken. This gives the typical value of the cross section to be that associated with the weak scale. The emergence of the weak scale from the relic abundance calculation is often cited as the WIMP miracle. Experimentally, there  are intense efforts to probe  dark matter cross sections \cite{planck2015, fermi} while  future experiments will further probe the annihilation rate for a wide range of dark matter masses \cite{hawc,cta}\footnote{For a complete list  of direct and indirect current and future experiments on dark matter searches see \cite{DMhub}.}. 

Given the ongoing experimental efforts, the time is ripe to critically examine the inputs that enter into the standard relic abundance calculation and explore alternatives where  dark matter is a thermal relic, yet
 the value of $\langle \sigma_{\rm th} v_{\rm rel} \rangle$ is different from the above quoted value. Of course, it could well be  that dark matter is non-thermal. Both avenues have received considerable attention (see for e.g. \cite{Barrow:1982ei, McDonald:1989jd, Kamionkowski:1990ni, Moroi:1994rs, Chung:1998rq, Hashimoto:1998mu, Moroi:1999zb, Giudice:2000ex, Allahverdi:2002nb, Endo:2006zj, Nakamura:2006uc, Acharya:2008bk, Allahverdi:2012wb, Allahverdi:2013tca, Allahverdi:2013noa, Co:2015pka, Aparicio:2015sda,KR}).

    A key input for the standard computation of  thermal dark matter abundance is that the energy density of the universe was dominated by 
radiation at the time that dark matter decoupled. One of the major successes of the hot big bang model is  nucleosynthesis (BBN), this requires that
the universe was radiation dominated at the time of production of the light nuclei, the corresponding temperature being $T_{\rm BBN} \simeq
{\rm 1 \spa MeV}$ (see \cite{bbn1,bbn2} for the related constraints on late time entropy production). But, the history of the universe prior to BBN  remains highly unconstrained. A modification of the history of the universe prior to nucleosynthesis, in particular a modified expansion rate of the universe can significantly alter the dark matter relic density computations.

 Scalar fields  with non-trivial evolution provide well motivated modifications of the pre-BBN history with interesting implications for dark matter relic abundances.   The simplest scenario with a modified cosmological history is  when an extra scalar species  is present \cite{decoupled}, which  does not couple directly to the Standard Model (SM) degrees of freedom or dark matter. The energy density of the decoupled scalar dilutes as  $\rho_\phi \propto a^{-(4+n)}$ with  $n>0$ (where $a(t)$ is the scale factor). This form implies that the scalar energy density dominates over radiation at early times. This in turn leads to a modification in the relic abundance calculations. Even this  simple model leads to novel features in the phenomenology --  dark matter  annihilation processes can continue to take place long after the decoupling of dark matter from the thermal bath. This was dubbed as  {\it relentless} dark matter, as it can be thought of as  dark matter trying to  get back into thermal equilibrium.  
 
More generally,   scalar-tensor (ST) theories of gravity arise  in extensions of the standard models of cosmology and particle physics. For example, in  higher dimensional  models,  additional scalar fields arise through the compactification of the extra dimensions and couple to the metric with gravitational strength.  Thus the gravitational interaction is mediated by both the metric and scalar fields so that
scalar-tensor gravity models represent a departure from standard General Relativity (GR). 
In string theory models of particle physics and cosmology,  new ingredients such as D-branes can also appear.  For such D-branes, the  longitudinal   fluctuations  are
identified  with  the  matter  fields  such  as  the  SM  and  DM  particles,  while  transverse
fluctuations  correspond  to  scalar  fields.  These scalars couple conformally and disformally to the matter living on the
brane  and  thus  can modify  the  cosmological  expansion  rate  felt  by  matter  and  hence the
standard predictions for the dark matter  relic abundance\footnote{Further  studies of the modifications to the relic abundances  in   conformally coupled scalar-tensor theories  have been discussed in \cite{Catena2, Gelmini, RG, WI, LMN,Pallis,Salati,AM,IC,MW2,MW3,MW}. } \cite{Catena,DJZ1,DJZ2}. 
   An attractive feature of the ST  case, is the possibility of tracker solutions of the scalar field, which ensure that the scalar field becomes inactive at the onset of BBN (and thereafter) as is required for its  successes. 

 One of the major successes of the inflationary paradigm is the explanation for the approximately scale invariant spectrum of the CMB. Given a model of inflation, the predictions for the scalar spectral tilt $(n_s)$  and the  tensor to scalar ratio $(r)$ are determined by the number of e-foldings, $N_k$, between horizon exit of the CMB modes and the end of inflation. Typically $N_k$ is taken to be in the range 50 to 60. It is important to keep in mind the inputs that go into deriving  this. The derivation makes use of two aspects of the physics of inflation: the condition for horizon exit of the CMB modes ($k = a_k H_k$; where $k$ is the wavenumber of the CMB modes, $a_k$ and $H_k$ are the scale factor and Hubble constant at the time of horizon exit) and the relationship between the strength of the scalar perturbations and the energy density at the time of horizon exit. Taking these inputs, the evolution of the energy
density from the time of horizon exit to the present epoch and entropy evolution from the end of  reheating to the present epoch are tracked. For the evolution, the post-inflationary history is taken to be the standard history associated with the hot big-bang model. This tracking then yields the equation that determines $N_k:$
\bel{nkstan}
N_k \approx 57 + {1 \over 4} \ln r - {1 \over 4} \left( 1 - 3 w_{\rm re} \right) N_{\rm re}\,,
\ee
where $r$ is the  tensor to scalar ratio and $w_{\rm re}, N_{\rm re}$ are the (effective) equation of state parameter and number of e-foldings during the reheating epoch. In the canonical reheating scenario, $w_{\rm re} \simeq 0$, but  more general cases can give $1/3 < w_{\rm re} <0$ \cite{ra,rp}. Given the uncertainties associated with reheating and the fact that the tensor mode has still not been detected, equation \pref{nkstan} does
not fix the value of $N_k$ exactly, but under the assumption that the reheating epoch is not highly extended it serves as the motivation for the 
usual range of 50 to 60 for $N_k$. The input that the post-inflationary history is the standard history associated with the hot big bang model is central to
the derivation of \pref{nkstan}, any modification of the post-inflationary history has an effect on the preferred range for $N_k$. As we enter the era of precision cosmology\footnote{It is expected that the next generation of CMB experiments \cite{next1,next2,next3} will bring $\Delta n_s$ at  the  $1-\sigma$ confidence level down to $0.002$, an improvement by a factor of three in comparison with Planck 2015 \cite{planck2015}. The  sensitivity for $r$ will come down to $10^{-3}$.}, it is very important to determine $N_k$ accurately,  so that theoretical predictions to confront the data can be made.  The models discussed above in the context of dark matter abundance have a non-standard cosmological history prior to BBN. The goal of this paper is to determine $N_k$  for these models and obtain the implications for inflationary predictions.

   The rest of the  paper is organised as follows. We review theories with scalars in pre-BBN cosmology motivated by dark matter considerations in section \ref{sectionsc}. In section \ref{modi}, we obtain the expressions for $N_{k}$ in these theories and in section \ref{implic} we discuss the implications  of our results for the predictions in the $(n_s,r)$ plane of various inflationary models. We  conclude in section \ref{conclusion}.

\section{Scalars in pre-BBN Cosmology}
\label{sectionsc}

\subsection{Extra  scalar species}

   The simplest scenario  to modify the cosmological history before the onset of BBN, is to consider a new  scalar  species $\phi$, which does not couple to matter, and has a non-trivial cosmological evolution  in the pre-BBN history. The defining property of a such decoupled scalar is that its energy
momentum tensor is conserved separately:  $ \nabla^{\mu} T^{\phi}_{\mu \nu} = 0$. In addition, various components of matter  have individually conserved stress tensors $\nabla^{\mu}T^{i}_{\mu \nu} = 0,$ where the index $i$ runs over the components of matter whose stress tensors are conserved. The consequence of  these conservation equations is that the various components of matter dilute independently and standard relationships between the number densities and the temperature\footnote{We will find that these relations do not hold in the Einstein frame when we discuss  scalars which have nontrivial couplings to the matter fields.} hold. To study alternative scenarios for dark matter production, the authors of  \cite{decoupled} considered cosmologies where the scalar energy density redshifts as
\bel{rsclae}
\rho_\phi \propto a^{-(4+n)} \ , \spa \spa \textrm{with} \spa \spa n > 0 \ .
\ee
The dilution can be expressed in terms of the temperature by making use of conservation of entropy
\bel{rscalet}
\rho_\phi(T) = \rho_\phi(T_r)  \left(\frac{g_{s}(T)}{g_{s}(T_r)}\right)^{(4+n)/3} \left(\frac{T}{T_r}\right)^{(4+n)} \ .
\ee 
where $g_{s}(T)$ is the effective number of degrees of freedom associated with the entropy density at temperature $T$, and $T_r$ is a reference temperature. 
The reference temperature is chosen such that the energy density of the scalar and radiation are equal at $T = T_r$.  With this, the deviation from standard cosmology is completely characterised by the two parameters $(n, T_r)$. 

The total energy density can then be written as
\bel{total}
\rho(T) = \rho_{\rm rad}(T) + \rho_\phi(T) =
\rho_{\rm rad}(T) \left[1 + \frac{g(T_r)}{g(T)} \left( \frac{g_{s}(T)}{g_{s}(T_r)} \right)^{(4+n)/3} \left(\frac{T}{T_r}\right)^{n} \right],
\ee
where $g(T)$ is the number of degrees of freedom associated with the energy density of the radiation at temperature $T$. The successes of BBN require that $T_r  >  (15.4)^{1/n} \spa \rm{MeV}$ \cite{decoupled}.  Typical values for $T_r$  considered in \cite{decoupled} are of the order of $\rm{GeV}$.
The relentless phenomenon occurs for $n \geq 2$ and $n\geq 4$ for $s$-wave and $p$-wave  annihilation, respectively \cite{decoupled}. 
We will use \eqref{total} in  section \ref{modi} to calculate the modification to the number of e-folds $N_k$ due to the pre-BBN non-standard cosmological evolution described above.


\subsection{Scalar-tensor theories }

 In this section, we describe the post-inflationary histories associated with a class of scalar-tensor theories, which are motivated from models with extra dimensions. It was shown in \cite{Bekenstein} that the most general physically consistent relation between two metrics $g_{\mu\nu}$ and $\tilde g_{\mu\nu}$ related by a single scalar field $\phi$ includes a conformal $C(\phi)$ as well as a disformal relation $D(\phi)$ as
 \be\label{gtilde}
\tilde g_{\mu\nu} = C(\phi) g_{\mu\nu} + D(\phi) \partial_\mu \phi \partial_\nu \phi\ .
\ee
The presence of the scalar modifies the expansion rate felt by any matter coupled to the ``disformal" metric $\tilde g_{\mu\nu}$, with interesting implications. The effect of this modified expansion rate for the thermal production of dark matter was recently studied in \cite{DJZ1,DJZ2}.  
We refer the reader to the literature \cite{DJZ1,DJZ2} for details of the general set-up and derivations. 

In these models, the total energy momentum tensor in the Einstein frame, which is defined with respect to the metric $g_{\mu\nu}$, is conserved: $ \nabla^{\mu} T^{tot}_{\mu \nu} =  \nabla^{\mu} (T^{\phi}_{\mu \nu}+  \sum_iT^{i}_{\mu \nu}) =0$, but the individual fluids are not, that is: $ \nabla^{\mu} T^{\phi}_{\mu \nu} = -\sum_i\nabla^{\mu} T^{i}_{\mu \nu} $. 
On the other hand, the energy momentum tensor in the Jordan, or disformal frame is conserved, $\tilde \nabla^{\mu}  \tilde T_{\mu\nu} =0 $, where $\tilde \nabla_{\mu}$ is the covariant derivative computed with respect to $\tilde g_{\mu\nu}$ and $ \tilde T_{\mu\nu} $ is the energy momentum tensor computed with respect to the disformal metric $\tilde g_{\mu\nu}$. 
The  energy densities and pressures in the two frames can be shown to be given by \cite{DJZ1,DJZ2}
\be\label{eqrhos}
\tilde \rho^i = C^{-2} \gamma^{-1} \rho^i \,, \qquad \tilde P^i = C^{-2}\gamma\, P^i \,,
\ee
where $\gamma = (1+\frac{D}{C}(\partial \phi)^2)^{-1/2}$ and  the equations of state in both frames are related by $\tilde \omega^{i} = \omega^{i}\, \gamma^2$.  In the pure conformal case,  $\gamma=1$ and  thus $\tilde \omega = \omega$. 
Finally, the Jordan and  Einstein frame  frame scale factors are related by 
\be\label{tildea}
\tilde a = C^{1/2} a \,.
\ee
The functions $C$ and  $D$ can in principle take any form so long as they satisfy the causality constraint $C(\phi)>0$,  $C(\phi) + 2X D(\phi) >0$, where $2X = (\partial \phi)^2$. Note also that in our conventions,  $C(\phi)$ is dimensionless, while  $D(\phi)$  has mass dimension $-4$.  In phenomenological studies, this is the only requirement on $C, D$ and the scalar field takes the standard action of a canonically normalised field.  On the other hand, in D-brane scalar-tensor theories,  the scalar action takes the  DBI form: ${\cal L}= M^4 C(\phi)^2\sqrt{1+2XD(\phi)/C(\phi)} + V(\phi) $, and the functions $C, D$ are  related as $M^4CD=1$. 
Moreover, $M$ is given by the tension of the D-brane, and thus to the string scale (see \cite{DJZ2}  and below). 
Finally, in this set-up, the inflaton  $\varphi$ may, or not, be disformally coupled to the scalar $\phi$, but it will anyhow have some  coupling and  will thus decay into both $\phi$ and  the matter fields coupled to it.

\section{E-foldings between Horizon exit of CMB modes and the end of inflation}

\label{modi}

   As discussed in the introduction, the number of e-foldings of the universe between the horizon exit of the CMB modes and the end of inflation $(N_k)$ is a central input for making predictions for the inhomogeneities in the CMB. The quantity is sensitive to  the post-inflationary history of the universe
and hence is expected to be altered with any modification of the pre-BBN history of the universe. In this section, we compute  $N_k$ for the theories
described in section \ref{sectionsc}. In spirit, the computations follow the standard procedure to determine $N_k$ (see for e.g. \cite{kam} for a recent discussion), but the modified post-inflationary histories in these theories lead to non-standard equations for $N_k$. 
\subsection{$N_k$ for decoupled scalars}\label{NDS}

   We start from the condition for horizon exit of modes during the inflationary epoch: $k  = a_k H_k $. Where $k$ is the comoving wavenumber of the mode,  $a_k$ and $H_k$ are the scale factor and Hubble constant at the time of horizon exit of the mode. We write the above as:
\bel{epochs}
   {k \over a_0} =  {a_{k} \over a_{\rm end} } \cdot {a_{\rm end} \over a_{\rm re} } \cdot{a_{\rm re} \over a_0 } \cdot H_k,
\ee
where $a_{\rm end}, a_{\rm re}$ are the scale factors at the end of inflation and end of the reheating epoch. $a_0$ is the scale factor today. We follow the  convention of tracking the evolution in terms of the number of e-foldings, \pref{epochs} then reads
\bel{logech}
    \ln \left(  {k \over a_0} \right) = - N_k - N_{\rm re} + \ln \left(  {a_{\rm re} \over a_0} \right) + \ln H_k.
\ee
Note that  if the entire post-inflationary history of the universe is known then $N_{\rm re}$ and $\ln \left(  {a_{\rm re} \big{/} a_0} \right)$, can be determined
by plugging in the values of the scale factor at  specific points of the evolution. The Hubble constant at the time of horizon exit is determined in terms
of the strength of scalar perturbations
\bel{hubk}
  H^2_k = { \rho_k  \over 3 M^2_{\rm pl}} = { \pi^2 \over 2 } M_{\rm pl}^2 A_s r,
\ee
where $A_s$  is the primordial scalar amplitude and $r$ the tensor to scalar ratio. Thus, the last term in \pref{logech} is known  modulo the value of $r$; the dependence on $r$ is logarithmic and  can be estimated without much uncertainly given a
model of inflation. Thus, given a post-inflationary history, \pref{logech} essentially determines $N_k$. While it is a perfectly valid approach to determine $N_k$ from \pref{logech}, it is possible to proceed further generally and determine what features of the post-inflationary history have an effect on $N_k$. To do this, we begin by a phenomenological parametrisation of the reheating epoch, taking the average equation of state during epoch to be $w_{\rm re}$.
We then have
\bel{repheno}
   N_{\rm re}  =  \ln \left( {a_{\rm re} \over a_{\rm end} } \right) =  {1 \over 3(1 + w_{\rm re}) } \ln \left( {\rho_{\rm end} \over \rho_{\rm re} } \right),
\ee
where $\rho_{\rm end}$ is the energy density of the universe at the end of inflation and $\rho_{\rm re}$ is the energy density at the end of reheating. We simplify the above by writing
\bel{rend}
  \ln ( \rho_{\rm end} )=    \ln \left( {\rho_{\rm end} \over \rho_k }\right) + \ln(\rho_k).
\ee
The first term involves the ratio of the energy densities at horizon exit and the end of inflation, it can be computed explicitly given the inflationary potential.
It typically is small as the energy density remains a constant during the inflationary epoch. The second term is related to the primordial scalar amplitude
and $r$ \pref{hubk}.  The right hand side of \pref{repheno} also involves  the energy density at the end of the reheating epoch $(\rho_{\rm re})$. We write this as
\bel{logre}
      \rho_{\rm re} = {\pi^{2} \over 30}g_{\rm re} \,T^4_{\rm re} \cdot \eta \,,
\ee
where $g_{\rm re}$ are the effective number of degrees of freedom at the end of reheating and $T_{\rm re}$ the reheat temperature of the Standard Model degrees of freedom. Usually, the Einstein frame energy density at the end of the reheating epoch is equated to the energy density of the radiation. As discussed in section \ref{sectionsc}, in the theories with a decoupled scalar proposed in \cite{decoupled} the decoupled scalar also carries a fraction of the energy density -- we  incorporate this in \pref{logre} by including the factor  $\eta$ (which is equal to the ratio of the total energy density and the energy density of the radiation at the end of reheating). Furthermore, we assume that there is no production of entropy\footnote{Entropy production can arise from modulus domination in the post-inflationary history \cite{kou, fq, banks}, the effect of this on inflationary predictions has been analysed in \cite{mod1, mod2, mod3, mod4}. } after the reheating epoch -- entropy conservation can then be used to relate the reheat temperature to the temperature of the CMB
\bel{entro}
  T_{\rm re} = \sqrt{ {43 \over 11 g_{\rm s, re}} } { \left( a_0 \over a_{\rm re} \right)} T_0.
\ee
Combining equations (\ref{repheno})-(\ref{entro}), we obtain: 
\bel{ncomb}
  { 3 \over 4} ( 1 + 3 w_{\rm re}) N_{\rm re}  = {1 \over 4} \ln \rho_k  - \ln \left( { a_0 T_0 \over a_{\rm re} } \right) +  {1 \over 4} \ln \left( {\rho_{\rm end} \over \rho_k }\right) + \beta - {1 \over 4} \ln \eta\,,
 \ee
 where
 \bel{beeta}
    \beta = - { 1 \over 4} \ln \left(   {\pi^{2} \over 30}g_{\rm re}  \right) - {1 \over 3} \ln \left( {43 \over 11 g_{\rm s, re}}\right) \,.
 \ee
Finally, we subtract \pref{ncomb}  from \pref{logech}  to eliminate the dependence on $ a_{\rm re} / a_{0}$. This yields an equation for $N_k$: 
\bel{nfinal}
    N_k = - {1 \over 4} ( 1 - 3 w_{\rm re}) N_{\rm re} + \ln H_k - {1 \over 4} \ln \rho_k - \ln \left( { k \over a_0 T_0} \right) - \beta + 
    {1 \over 4} \ln \left( {\rho_k \over \rho_{\rm end}} \right) + {1 \over 4} \ln \eta\,.
\ee
Let us discuss each term in the right hand side. The first term depends on the parametrisation of reheating $(N_{\rm re}, w_{\rm_{re}})$. Given the 
usual uncertainties associated with it, this term is the major source of the 
uncertainty in $N_k$. The second and third terms are determined in terms
in terms $A_s$ and $r$ \pref{hubk}. In the fourth term, $k/a_0$ is the pivot scale for CMB observations and $T_0 = 2.73\spa K$. The fifth term depends on the number of effective degrees of freedom at the time of reheating; $\beta \propto \ln \left( g_{\rm re}^{1/4} \big{/} g_{\rm s, re}^{1/3} \right)$, hence it gives a  small contribution.  The sixth term involves the ratio of the energy densities at the time of horizon exit and the end of inflation. It is typically small as the energy density does not change
much during inflation. The last term involves the parameter $\eta$ introduced in \pref{logre}. In conventional cosmological scenarios, this term vanishes as the entire energy density of the universe is in the form of radiation after reheating. 
Plugging in the numerical factors we find\footnote{We take $g_{\rm re} \approx g_{\rm s, re} \approx 100$ and $ {\rho_{\rm end} \big{/} \rho_k } \approx 1$.}
\bel{nnumber}
   N_k \approx  57 -  {1 \over 4} ( 1 - 3 w_{\rm re}) N_{\rm re} + {1 \over 4} \ln r + {1 \over 4} \ln \eta.
\ee
As discussed above, in conventional cosmological scenarios $\eta=1$.  The term involving the reheating parameters is typically negative, simulations of   various  scenarios for reheating give $w_{\rm re} < 1/3$  (see for e.g. \cite{ra,rp}). The term involving $r$ is manifestly negative. Given this, equation \pref{nnumber}  gives motivation for the usual range of $50$ to $60$ for $N_k$ in conventional cosmologies, (i.e. $\eta=1$). We can read off  the expression for $\eta$ (in terms of $n$ and $T_{r}$) from the expression for the total energy in the post-inflationary epoch\footnote{ Assuming$\frac{g(T_r)}{g(T_{\rm re})} \left( \frac{g_{s}(T_{\rm re})}{g_{s}(T_r)} \right)^{(4+n)/3} \sim {\cal O}(1)$. Note that in general, this factor is greater than one, hence when relevant, it will lead to a further enhancement of the effect}. Equation \eqref{total} gives
\bel{decs}
   \eta \approx \frac{g(T_r)}{g(T_{\rm re})} \left( \frac{g_{s}(T_{\rm re})}{g_{s}(T_r)} \right)^{(4+n)/3} \left(\frac{T_{\rm re} }{T_r}\right)^{n}.
\ee
\smallskip
\noindent Recall that the $\eta$ contribution to $N_k$ is:  $ \delta N_k \equiv 1/4 \ln \eta$ \pref{nfinal}. 

\begin{figure}[!htb!]
\begin{center}
\includegraphics[width=0.6\textwidth]{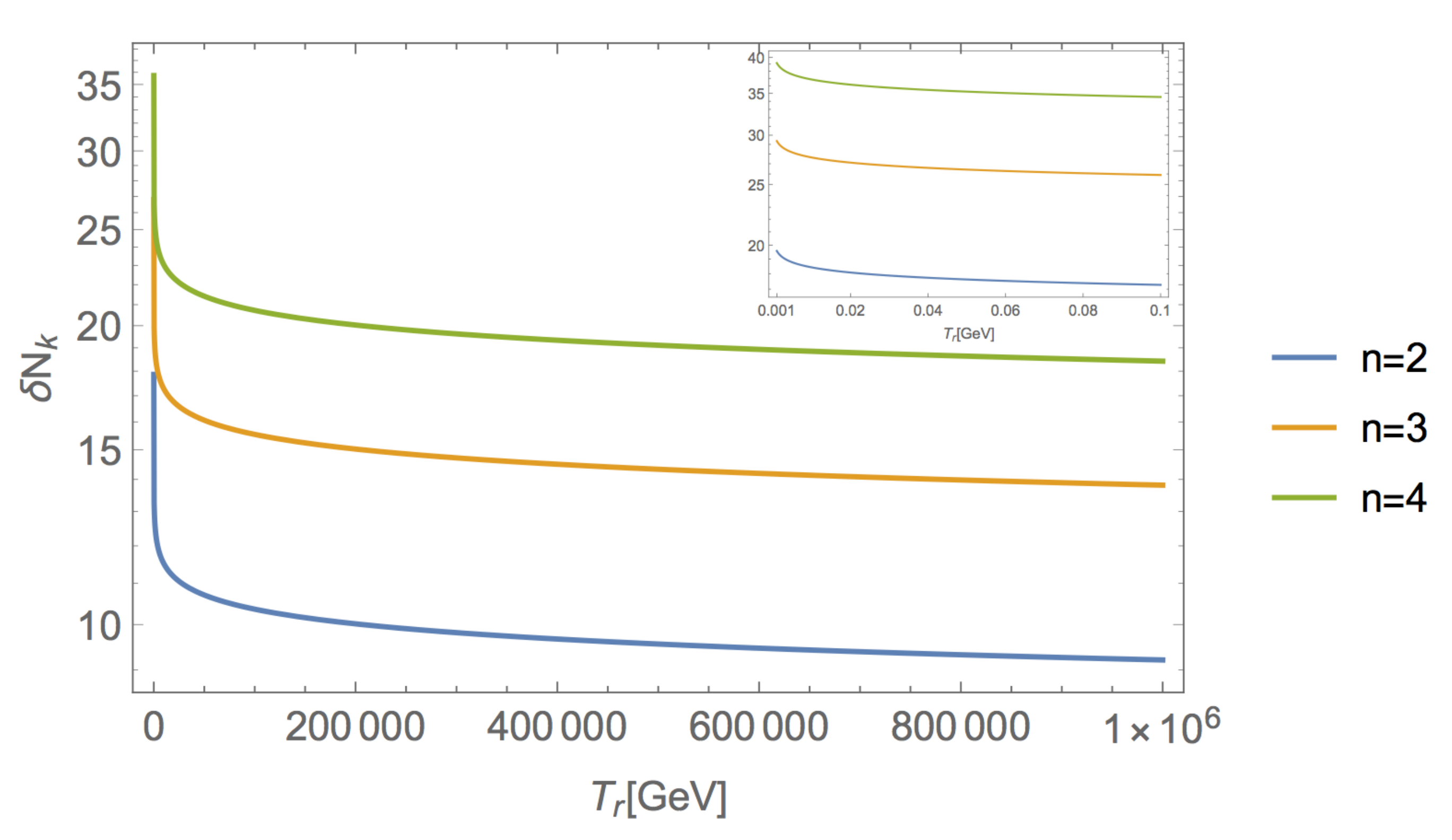}
\caption{$\delta N_{k}$ %
as a function of $T_r$, for different values of $n$ with $T_{\rm re} = 10^{14} \, \rm{GeV}$.}
 \label{f1}
\end{center}
\end{figure}

 As was shown in \cite{decoupled}, the largest effect in the dark matter decoupling  is obtained for smaller $T_r$'s, so long as the BBN predictions are not affected, which gave $T_r\simeq 10$ MeV.   From \eqref{decs} we see that the largest effect in $\delta N_k$ for a given reheating temperature, will precisely be given for the smallest $T_r$'s  as  shown in Fig.~\ref{f1}.

Taking  for example a reheating temperature of $T_{\rm re} = 10^{14}~\rm{GeV}$,   for  $n=2$ and  $ T_{r} = 1~\rm{GeV} $ one obtains $ \delta N_k \approx 16$.  For $n=2,  \spa T_{r} = 10^6~\rm{GeV} $,  $ \delta N_k \approx 9$. Since $n$ appears in the exponent in \pref{decs}, higher values of $n$ can lead to much larger shifts in  $N_k$.  For $n=4$ (recall that this is the critical value of $n$ for relentless dark matter if dark matter annihilation is a $p$-wave process \cite{decoupled}) with  $T_{r} = 1~\rm{GeV} $ yields $ \delta N_k \approx 32$, while $T_{r} = 10^6~\rm{GeV} $  yields $ \delta N_k \approx 18$. 
Note that since $T_{\rm re}\gg T_r$,  the change in $\delta N_k$ is larger for smaller values of $T_r$ (see inset in Fig.~\ref{f1}). For example, for a factor $m\sim {\cal O}(1)$  change in $T_{\rm r}$, the change in $\delta N_k\sim \frac{n}{4}\log m$, which gives an $ {\cal O}(1)$ change (see inset in Fig.~\ref{f1}).

We will discuss the implications of our results for confronting inflationary models with data  in section \ref{implic},  here would like to emphasise that shifts in the value of $N_k$ obtained above can have a significant effect on the $(n_s, r)$   predictions of any inflationary model. Also, note that since $\eta$ is the ratio of the total energy density and the energy density in radiation at the end of reheating, $\eta > 1$; hence the effect of the last term in \pref{nnumber} is to increase $N_k$. The positivity of
the shift is natural as in the epoch in which the scalar dominates the energy density in the post-inflationary history the energy density dilutes as 
$a^{-(4+n)}$, i.e faster than radiation. This is similar to a kination phase \cite{kin1,kin2, kin3, kin4}, which is known to make a positive contribution to $N_{k}$ (see for e.g. \cite{dimo} for a recent discussion). Of course, the key difference is that this phase occurs after reheating.

\subsection{$N_k$ for coupled scalars}

\label{cop}

In this section, we  derive the expression for the  number of e-foldings between  horizon exit of CMB modes and the end of inflation ($N_k$) for the coupled case where the scalar arises from a scalar-tensor theory and thus couples both conformal and disformally to matter.  The crucial  difference from the discussion in the previous section is  that  the standard relations for energy and entropy conservation now hold in the Jordan frame\footnote{As mentioned earlier, we shall denote all Jordan frame quantities by a tilde superscript to distinguish them frame Einstein frame quantities.}. The  analogous expression for  
 \eqref{logech} in the present case is given by 
\bel{logechC}
    \ln \left(  {k \over a_0} \right) = - N_k - \tilde N_{\rm re} + \ln \left(  {\tilde a_{\rm re} \over  a_0} \right) + \ln H_k\,,
\ee
where 
\be\label{tN}
\tilde N_{\rm re}=  \ln \frac{\tilde a_{\rm re}}{\tilde a_{\rm end}} = N_{\rm re}  +\frac{1}{2} \ln \frac{C_{\rm re}}{C_{\rm end}}\,,
\ee
and since the scalar is  not active at the end of inflation, we can take $\tilde a_{\rm end}=a_{\rm end}$ (that is, $C_{\rm end}=1$). Moreover, since  standard GR evolution must be reached at the onset of BBN to avoid spoiling its predictions, we also have  $\tilde a_0 = a_0$. 
Now, using energy conservation in the Jordan frame,  we  can write 
\bel{reC}
   \tilde N_{\rm re}  =  \ln \left( {\tilde a_{\rm re} \over  a_{\rm end} } \right) =  {1 \over 3(1 +  w_{\rm end}) } \ln{\rho_{\rm end}  } -   {1 \over 3(1 + \tilde w_{\rm re}) } \ln{ \tilde \rho_{\rm re} }. 
\ee
We gave an expression for $\rho_{\rm end}$ in \eqref{rend} in terms of $\rho_k$. On the other hand, in the Jordan frame we have \cite{DJZ1,DJZ2} 
\bel{logreC}
    \tilde   \rho_{\rm re} = {\pi^{2} \over 30}g_{\rm re} \tilde T^4_{\rm re} \,,
\ee
where $g_{\rm re}$ is the effective number of degrees of freedom (which in principle depend on $\tilde T$) at the end of reheating and $\tilde T_{\rm re}$ is the the Jordan frame reheat temperature. 
Furthermore,  assuming as before that there is no production of entropy after the reheating epoch -- entropy conservation in the Jordan frame can  be used to relate the reheat temperature to the temperature of the CMB:
\be
\tilde  T_{\rm re} = \left( \frac{43}{11 g_{\rm s,re}}\right)^{1/3} \left(\frac{a_0}{\tilde a_{\rm re}}\right) \tilde T_0\,.
\ee
Using these equations, we arrive at an expression for $N_k$ in terms of $\tilde N_{\rm re}$:
\be
N_k = -\frac{1}{4}(1-3\,\tilde \omega_{\rm re}) \tilde N_{\rm re}  +\ln H_k - \beta - \ln\left( \frac{k}{a_0 T_0}\right) -
 \frac{(1+\tilde \omega_{\rm re})}{4(1+ \omega_{\rm end})} \left[\ln \left(\frac{\rho_{\rm end}}{\rho_k} \right) +\ln \rho_k\right]\,.
\ee
Using \eqref{tN}, and considering that $\tilde \omega_{\rm re} \sim \omega_{\rm end}$, we arrive at 
\be
N_k = -\frac{1}{4}(1-3\tilde \omega_{\rm re})N_{\rm re}  +\ln H_k - \beta - \ln\left( \frac{k}{a_0 T_0}\right) -
 \frac{1}{4} \ln \left(\frac{\rho_{\rm end}}{\rho_k} \right) +  \frac{1}{4} \ln \rho_k
  -\frac{1}{8}(1-3\,\tilde \omega_{\rm re})  \ln C_{\rm re}   \,.
\ee
In general $C(\phi)$ depends on time (or temperature). However, as studied in \cite{DJZ2}, a pure disformal term with $C=C_0=const.$  gives a non-trivial enhancement of the expansion rate, and thus an earlier dark matter freeze-out than in the standard case. This  causes a relentlessness  effect, similar to the decoupled case discussed in the previous section.
Therefore a constant $C_0$  is the most interesting example. Furthermore,  we can also assume a Jordan frame equation of state $\tilde \omega_{re} $ to be roughly constant. 

In the case of a D-brane scalar tensor model \cite{DJZ2}, the disformal and conformal functions arise from the DBI action as discussed above. The scale $M$ introduced there is related to the tension of a D3-brane, which is given by the string scale as $M=M_s (2\pi g_s^{-1})^{1/4}$. Defining the  scale $\tilde M = M C_0^{1/4}$, the interesting cases for the modified dark matter thermal scenarios occurred for $\tilde M \in (10-300)  \, \,{\rm GeV}$ in \cite{DJZ2}. Considering a typical string scale, which arises in string theory models of inflation, $M_s\sim (10^{14}-10^{16})\,\,{\rm GeV}$ (with $g_s\sim 0.1$) allows us to find the value of $C_0$ and thus the modification to $N_k$. 
The net effect is  a positive modification to $N_k$, hence  increasing the required number of efolds by about $\delta N_k \sim 15$ (for $\tilde \omega_{\rm re} \sim 0$), and thus similar to the decoupled example with e.g.~$n=2$, $T_{r} = 1~\rm{GeV} $ and  $T_{\rm re} = 10^{14}~\rm{GeV}$. 
In terms of a D3-brane picture,  this can be interpreted as the D-brane moving at the tip of a warped throat.  

The case $C_0=1$ on the other hand, would  correspond to an unwarped geometry. In this case,  $\tilde M=M$ and therefore an effect in the thermal dark matter production  is  tied to the string scale. In this case the relevant values of $M$ for a relentless dark matter  effect, require extremely low string scales \cite{DJZ2}. 
In the phenomenological case on the other hand, the values of $C_0$ and $D_0$ are in principle unconstrained and therefore can again give large positive corrections to $\delta N_k$. 

%


\section{Implications for Inflationary Predictions}
\label{implic}

In this section, we discuss the implications of the shifts in $N_{k}$ derived in section \ref{modi} for the predictions of inflationary models in the $(n_s, r)$ plane. Inflationary models fall into universality classes  characterised by the relationship between $n_s, r$ and $N_{k}$ \cite{Roest:2013fha}. It is natural to use this classification to study  how the shift in $N_{k}$ can affect various models. Class I models obey the relations
\bel{class1}
     n_{s} \simeq 1 + { 2 \lambda \over N_k} , \spa \spa \spa r \sim {1 \over N_k^{-2 \lambda}} \spa \spa \spa \lambda <  -\hf \,,
\ee
where $\lambda$ is a constant which depends on the inflationary model. Note  that a increase $N_k$ implies that models with higher values of $\eta$ are preferred (similarly higher values of $N_k$ imply higher values of $\eta$). The prototypical
models for Class I are hilltop  \cite{Boubekeur:2005zm} and Starobinsky \cite{Starobinsky:1980te}. Amongst the models obtained from string theory, 
it includes the fibre inflation model \cite{Cicoli:2008gp, shanta, Cicoli:2017axo}. The Starobinsky model has
\bel{class1}
     n_{s} \simeq 1 - { 2  \over N} , \spa \spa \spa r \simeq {12 \over N^{2 }} \,,
\ee
which corresponds to $\lambda = -1$.  For these models lowering of $N_k$  (as found in \cite{mod1})  can be interesting as this pushes up the value of $r$, the increase in $N_k$ found for the models makes the size of $r$ even smaller.

   Class II models  are characterised by
\bel{class2}
   n_{s} \simeq 1 - { 2\lambda +1  \over N} , \spa \spa \spa r \simeq {16 \lambda \over N}\,,
\ee
where again $\lambda$ is a constant  depending on the inflationary model.
These are large field models.  The prototypes for these are the monomial potentials: $V(\varphi) = {\hf} m^{4- \alpha} \varphi^{\alpha}$ ($\alpha = 2$ corresponds to the famous $m^2 \varphi^2$ model \cite{Linde:1983gd, Belinsky:1985zd, piran}, $\alpha = 2/3, 1$ arise in the  axion monodromy examples \cite{mac1, mac2}). The predictions in the 
$(n_s, r)$ plane are 
\bel{ns}
  n_{s} = 1 - {\alpha +2 \over 2N}; \spa  r=  { 4 \alpha \over N} \,.
\ee
For the $m^{2} \varphi^{2}$ model, $N_{k}$ in the range of 
50 to 60, gives $n_s$ in the range 0.960 to 0.966 and $r$ in the range $0.13$ to $0.16$. Recall that 
 {\sc{Planck}} 2015 (TT+TE+EE) + low P + lensing for the $\Lambda$CDM + $r$ model gives $n_s = 0.9688   \pm 0.0061$ and $r < 0.114$ at $1-\sigma$ \cite{planck2015}). The major tension with data arises from the high value of the prediction for $r$. This can be ameliorated with an increase in the value in $N_k$. For e.g. $N_{k} = 72$ gives $r=0.11$ and $n_s = 0.972$, which is within the Planck 2015 $1-\sigma$
 range for $n_s$ and $r$. In scenarios for which the reheating epoch is not extended, this corresponds to $\delta N_{k} \approx +16$. Such a shift is possible for a large range of the model parameters both the decoupled and coupled scalar cases; particularly in the case of relentless dark matter  
 and the  D-brane scalar-tensor model \cite{DJZ2}. We note that although the shift brings both $n_s$ and $r$ within their $1-\sigma$ values, the model only enters the 
the $2-\sigma$ region in the marginalised joint distribution in the $(n_s, r)$ plane\footnote{With the Planck 2015 data, it impossible for the model to enter $1-\sigma$ region in the joint distribution, for the model to predict $r$ values in the necessary range  requires $N >100$, but in this regime the predictions for $n_s$ are incompatible with the data.}. But overall the shift in the predictions is in the direction favoured by the present data. For the linear potential $(\alpha =1)$,  $N_k$ in the range of 50 to 60 gives $n_s$ in the range $0.980$ to $0.983$ and $r$ in the range $0.080$ to $0.066$. In this case the prediction for $n_s$ is on the higher side, and increase in   $N_{k}$ further increases the tension in the $n_s$ prediction. The situation for $\alpha = 2/3$ is similar.

Another interesting class of models is that of natural inflation \cite{Freese:1990rb}. In this case the dependence on $N$ of the cosmological parameters ($n_s,r$) is non-perturbative with: 
\be
n_s=   1+ \frac{1}{f^2}\frac{(e^{N_k /f^2} +1)}{(e^{N_k /f^2} -1)} \,,  \qquad \quad r = \frac{8}{f^2} \frac{1}{e^{N_k /f^2}-1}\,,
\ee 
 where $f$ is the axion decay constant (in Planck units).     
  Therefore, a larger value of $N_k$ can bring this model back into the allowed region of the parameters. For example, for $f=7M_P, N_k=60$ we have $(n_s, r) = (0.962, 0.07)$, while $f=7M_P, N_k=75$ gives  $(n_s, r) = (0.968, 0.05)$. We show the model predictions as a function of $N_k$ in Fig. 2
\begin{figure}[!htb!]
\begin{center}
\includegraphics[width=0.55\textwidth]{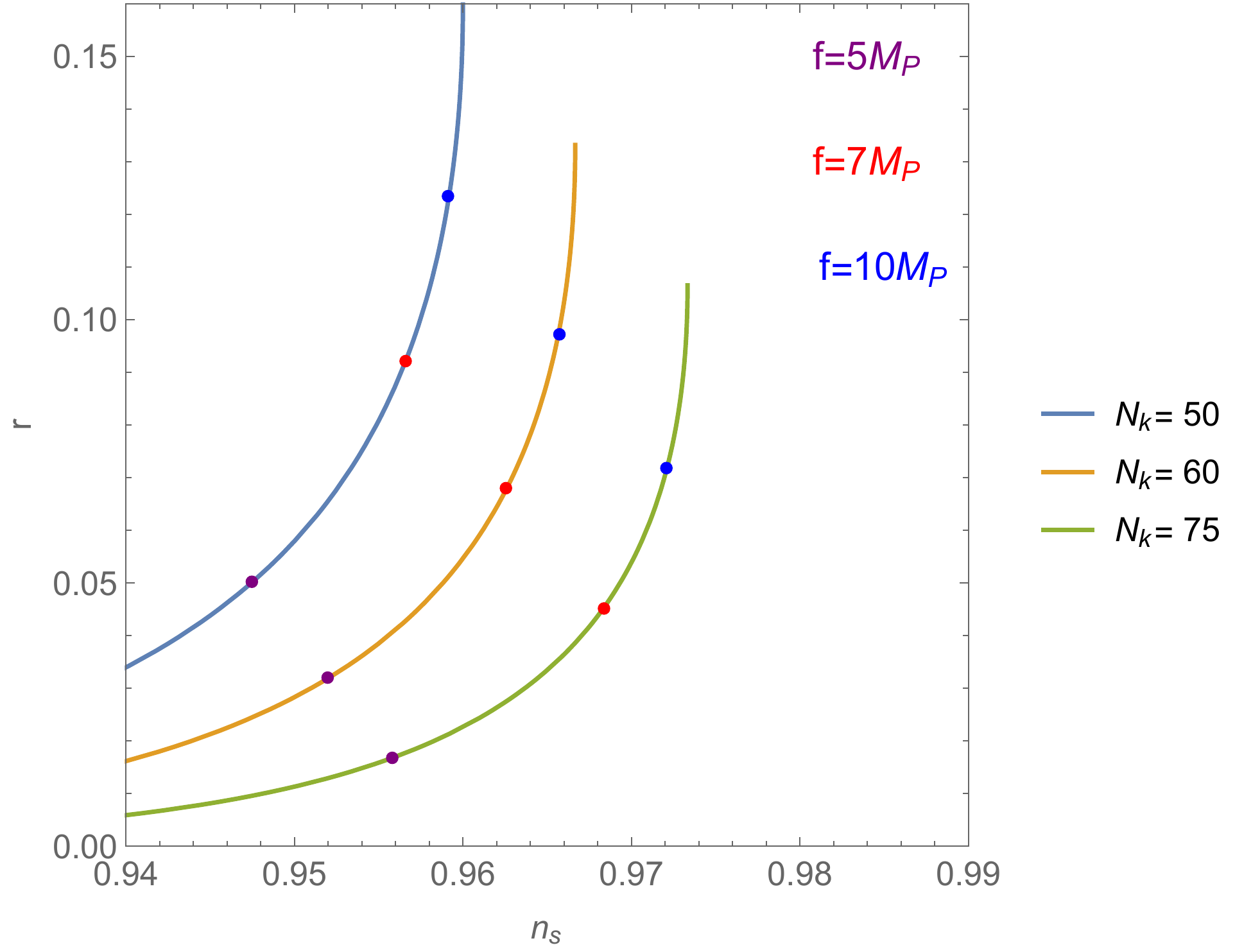}
\caption{Predictions of Natural Inflation in the $(n_s,r)$ plane.}
 \label{f2}
\end{center}
\end{figure}  

\section{Conclusions}\label{conclusion}

 Scalar fields are ubiquitous in theories beyond standard models of cosmology and particle physics. During the early pre-BNN cosmological history, these can appear as an extra species, decoupled from matter, or in scalar-tensor theories, where the gravitational interaction is mediated by both the metric and scalar field, coupling  to matter conformal and disformally. 

These scalars evolve during the pre-BNN epoch, giving rise to interesting implications for dark matter relic abundances \cite{Catena,decoupled,DJZ1,DJZ2}.  The modified post-inflationary history
 implies a shift in  the number of e-foldings between the horizon exit of CMB modes and the end of inflation,  $N_k$. At the same time, this implies a change in the inflationary parameters  $(n_s,r)$. We have computed this shift for
 theories with decoupled scalars and scalar-tensor theories and examined the effect that this can have on the predictions in the $(n_s,r)$ plane for inflationary models. In particular we considered the models discussed recently in \cite{decoupled} when an extra scalar species modifies the DM  relic abundance and the disformal models arising from D-brane scalar-tensor theories discusses in \cite{DJZ1,DJZ2}, where the scalar couples conformal and disformally to matter,  modifying the expansion rate and thus the DM relic abundances.  

For decoupled scalars, we found that the shift in $N_k$ is positive definite. The case of relentless DM  gives the largest effect with  $\delta N_k \approx 15$ or larger, with a reheating temperature of around $T_{\rm re}\sim 10^{14}$GeV (see section \ref{NDS}). 
For the scalar-tensor case, we derived $\delta N_k$ for  a  disformal  enhancement of the expansion rate as studied  in \cite{DJZ2}. This corresponds to constant conformal and disformal functions $C=C_0$, $D_0= \tilde M^{-4} =(C_0M^4)^{-1}$. For the values of $\tilde M$ and $C_0$ relevant for the modifications of the expansion rate and thus relic DM abundance  \cite{DJZ2}, we found a  large positive effect with   $\delta N_k \approx 15$ (see section \ref{cop}).

We used this to study the implications for the predictions of $(n_s, r)$ for some interesting  inflationary models.  For example, the increase in $N_k$  implies the possibility of a decrease in the prediction for $r$ for the $m^2 \varphi^2$ model, which reduces the tension with data.  
The change in $N_k$ has also an important effect for natural inflation, by decreasing $r$ and moving $n_s$ towards the allowed values of Planck. 
Analysis in the spirit of this work will become more and more important as we probe the CMB even more minutely.

\section*{\Large{Acknowledgements}}
 We would like to thank the organisers of the workshop  on Post-Inflationary String Cosmology, Bologna  2017 for the simulating  environment which led this collaboration. AM is partially supported by a Ramanujan Fellowship, DST, Government of India. IZ is partially supported by STFC grant ST/P00055X/1.

\bibliography{refs}

\bibliographystyle{utphys}

\end{document}